\documentclass[twocolumn,floatfix,superscriptaddress,amsmath,showpacs,showkeys,aps,pre]{revtex4-2}

\usepackage{epsfig}
\usepackage{bm}
\usepackage{color}
\usepackage{amsmath}
\usepackage{amssymb}
\usepackage{graphicx}
\usepackage{physics}
\usepackage{xr}
\usepackage{soul}
\usepackage{longtable}
\usepackage{verbatim}
\usepackage[OT1]{fontenc}
\usepackage[normalem]{ulem}
\usepackage{blkarray}

\newcommand{\mm}[1]{\langle #1 \rangle}

\begin{document}

\title{Mean-field solution of the neural dynamics in a Greenberg-Hastings model with excitatory and inhibitory units}

\author{Joaquin Almeira}

\affiliation{Instituto de F\'isica Enrique Gaviola (IFEG-CONICET) Ciudad Universitaria, 5000 C\'ordoba, Argentina}

\author{Tomas S. Grigera}
\affiliation{Instituto de F\'isica de L\'iquidos y Sistemas Biol\'ogicos (IFLYSIB), CONICET y Universidad Nacional de La Plata, La Plata, Argentina}
\affiliation{Departamento de F\'isica, Facultad de Ciencias Exactas, Universidad Nacional de La Plata, La Plata, Argentina}
\affiliation{Istituto dei Sistemi Complessi, Consiglio Nazionale delle Ricerche, via dei Taurini 19, 00185 Rome, Italy}
 \affiliation{Consejo Nacional de Investigaciones Cient\'{\i}fcas y Tecnol\'ogicas (CONICET), Buenos Aires, Argentina}

 \author{Daniel A. Martin}
 \affiliation{Instituto de Ciencias F\'isicas (ICIFI-CONICET), Center for Complex Systems and Brain Sciences (CEMSC3), Escuela de Ciencia y Tecnolog\'ia, Universidad Nacional de Gral. San Mart\'in, Campus Miguelete, San Mart\'in, Buenos Aires, Argentina}

\author{Dante R. Chialvo}
\affiliation{Instituto de Ciencias F\'isicas (ICIFI-CONICET), Center for Complex Systems and Brain Sciences (CEMSC3), Escuela de Ciencia y Tecnolog\'ia, Universidad Nacional de Gral. San Mart\'in, Campus Miguelete, San Mart\'in, Buenos Aires, Argentina}
\affiliation{Consejo Nacional de Investigaciones Cient\'{\i}fcas y Tecnol\'ogicas (CONICET), Buenos Aires, Argentina}

\author{Sergio A.\ Cannas}
\affiliation{Instituto de F\'isica Enrique Gaviola (IFEG-CONICET) Ciudad Universitaria, 5000 C\'ordoba, Argentina}

\affiliation{Facultad de Matem\'atica Astronom\'ia F\'isica y Computaci\'on, Universidad Nacional de C\'ordoba.}
\affiliation{Consejo Nacional de Investigaciones Cient\'{\i}fcas y Tecnol\'ogicas (CONICET), Buenos Aires, Argentina}

\date{\today}

\begin{abstract}
We present a mean field solution of the dynamics of a Greenberg-Hastings neural network with both excitatory and inhibitory units. We analyse the dynamical phase transitions that appear in the stationary state as the model parameters are varied. Analytical solutions are compared with numerical simulations of the microscopic model defined on a fully connected network. We found that the stationary state of this system exhibits a first order dynamical phase transition (with the associated hysteresis) when the fraction of inhibitory units $f<  f_t \leq 1/2$, even for a finite system. In finite systems, when $f > f_t$ the first order transition is replaced by a pseudo critical one, namely a continuous crossover between regions of low and high activity that resembles the finite size behaviour of a continuous phase transition order parameter. However, in the thermodynamic limit, we found that $f_t\to 1/2$ and the activity for $f\geq f_t$ becomes negligible for any value of $T>0$, while the first order transition for $f<f_t$ remains.
\end{abstract}

 \maketitle

\section{Introduction}

The theory of phase transitions in systems under thermodynamical equilibrium is a well established statistical mechanic's formalism. On the contrary,  the analogous phenomena in non-equilibrium systems present a much more richer scenario that appears to be more complex to formalise, so a general theory is still lacking. In particular, critical phenomena and phase transitions in non-equilibrium systems seem to be more sensitive to the microscopic details ({\it i.e.}, the dynamic rules) than the equilibrium counterpart. In other words, universality seems to be more restricted in non-equilibrium systems than in thermodynamical ones, one of the reasons behind the larger variety of observed behaviours. Hence, the path towards the development of a general theory of the topic at the present strongly relies on comparing the behaviour of as many different models as possible, with the as deep as possible understanding of them.

In this work, we considered a basic model of neural dynamics based on 
Greenberg-Hastings (GH) cellular automaton of excitable media\cite{GH}, which has shown recently a very rich phase transitions scenario\cite{Haimovici}--\cite{Almeira} and provided support to the so-called ``brain criticality  hypothesis''\cite{Beggs}--\cite{Mora}. But besides its interest in neuroscience,  the complexity exhibited by the model highlights its importance {\it per se} to the field of non-equilibrium phase transitions theory. Up to now, all the studies of this model were based on numerical simulations. In this work, we present a mean field solution of the dynamics of the model. Mean field solutions are always of theoretical interest, not only because analytical expressions are a powerful tool that allow a deeper understanding of the problem, but also because they provide us with some limit cases to check the results of models defined on more realistic networks.

In the model here considered a  three state dynamical variable $x_i = 0,\,1,\,2$ is associated to each node $i$ of some network $\Omega$,  representing the following dynamical states: quiescent ($x_i=0$), excited ($x_i=1$), and refractory ($x_i=2$). Also to each node $i$ there is associated a quenched random variable $\epsilon_i=\pm 1$, representing an excitatory or inhibitory unit respectively. There is a fraction $f$ of inhibitory units in the network, namely, each node has a probability  $f$  to be inhibitory ($\epsilon=-1$) and $1-f$ to be excitatory ($\epsilon=+1$). The model follows a parallel dynamics in a discrete time $t$, where the transition probabilities  for the \textit{i}-th site are given by

\begin{eqnarray}\label{eq:probGH}
    P_i(0\rightarrow1) &=& 
    1-(1-r_1) A\left(\{ x_j\} \right)\nonumber \\
    P_i(1\rightarrow2) &=& 1\nonumber\\
    P_i(2\rightarrow0) &=& r_2.
\end{eqnarray}
\noindent  $r_1$ is the probability of spontaneous activation, $r_2$ is related to the average refractory period (with an average duration of $1/r_2$), and the activation function $A\left(\{ x_j\} \right)$ is given by

\begin{equation}
    A\ =  \qty[1-\Theta\qty(\sum_{j\in \Omega_i} W_{ij}\epsilon_j\delta(x_j,1)-T)]
\end{equation}

\noindent where  $\Theta$ is the Heaviside function and $\delta(x,y)$ is the Kroenecker delta function. $W_{ij}$ is a symmetric synaptic matrix, $\Omega_{i}$ is the set of nearest neighbours of the node $i$ and $T$ is the activation threshold, assumed to be the same for all neurons. 

 A first version of this model with purely excitatory neurons $f=0$ was introduced by Haimovici {\it et al.}\cite{Haimovici}. The synaptic matrix $W_{ij}$ in this case was constructed based on an empirical structure of neuroanatomical connections of about $N \sim 1000$ nodes\cite{Hagmann}. The consequent analysis of cluster statistics provided the first evidence of critical behaviour in this model\cite{Haimovici}. Further evidence about criticality was obtained through the implementation of the model on small world networks (Watts-Strogatz model) with arbitrary size $N$ (still with $f=0$). Such implementation allowed to perform a finite size scaling analysis\cite{Zarepour}. Moreover,  a very rich dynamical phase diagram emerged as the topological properties of the network (namely, the average degree $\langle k \rangle$ and the rewiring probability $\pi$) were varied\cite{Zarepour,Sanchez}. In this case, the non-null  synaptic weights (given by the adjacency matrix of a Watts-Strogatz network) were quenched random variables with an  exponential distribution

\begin{equation}\label{eq:hagmann}
    p(W_{ij}=w) = \beta e^{-\beta w}. 
\end{equation}

 Furthermore, the inclusion of inhibitory units $f\neq 0$ to the previous model revealed the possibility of tricritical behaviour in the $(f,T)$ space, depending on the topology of the network\cite{Almeira}.

 In this work, we consider the implementation of the model on a fully connected network where the probability distribution for the non-null weights is given by Eq.(\ref{eq:hagmann}). This case corresponds to the large degree limit $\langle k\rangle\to N\to\infty$ of the (properly normalized) GH model on small world networks previously considered\cite{Almeira}. 
We obtained a set of differential equations for the average population densities of excited sites and focused our analysis on the stationary solutions of those equations, both for finite size systems and in the infinite size limit.

 The paper is organised as follows. In section \ref{model} we present the model and describe the stochastic technique used to estimate the mean field dynamical equations. In section \ref{results} we analyse the stationary solutions of the mean-field equations in the $(f,T)$ space and compare them with some numerical simulations of the fully connected model. A general discussion of the results is presented in section \ref{conclusion}.

\section{Model and methods}
\label{model}

The fully connected model here considered is defined by the synaptic matrix
\begin{equation}
       %\textcolor{red}{ W_{ij} = W_{ji} \sout{= \frac{1}{N} w_{ij}} }\\
        w_{ij} = w_{ji} = \frac{1}{N} W_{ij}
\end{equation}

\noindent for all pairs of nodes $(i,j)$ ($\Omega_i = \Omega$) with $w_{ii} = 0$ (no self-connections) and $p(W_{ij})$ given by Eq.(\ref{eq:hagmann}). In order to be consistent with the large $\mm{k}$ limit of the small-world based models in Refs.\cite{Zarepour}-\cite{Almeira} we also use  the values $r_1 = 10^{-3}$ and $r_2 = 0.3$ and $\beta=12.5$.

For long enough time scales it is expected the evolution of this model to be well described as a continuous time Markov stochastic process in the variables 
 $\va*{x}(t) = (x_1(t),\dotsc,x_N(t))$, with appropriated chosen transition rates between states. We will {\it assume} the following transition rates for $x_i(t)$
\begin{equation}\label{eq:transitionrates}
\begin{split}
    U_i(0\rightarrow1) &= \mu_1 + \alpha\,\Theta\qty(\sum_{j=1}^N w_{ij}\epsilon_j\delta(x_j(t),1)-T)\\
    U_i(1\rightarrow2) &= \mu_3\\
    U_i(2\rightarrow0) &= \mu_2
\end{split}
\end{equation}
\noindent where $\mu_1$, $\mu_2$, $\mu_3$, and $\alpha$ are constants. The consistency of this ansatz will be checked later and the results compared with numerical simulations.  While Eqs.(\ref{eq:transitionrates}) allow in principle to construct a master equation for the probability $P(\va*{x},t)$, it would be analytically intractable. Thus, it is convenient to work with macroscopic variables like the number of sites in every state, such as $n_q^E = \sum_{i\in \text{exc}} \delta(x_i,0)$, $n_e^E = \sum_{i\in \text{exc}} \delta(x_i,1)$, and $n_r^E = \sum_{i\in \text{exc}} \delta(x_i,2)$ for excitatory sites (analogously for inhibitory ones). A master equation for the probability distribution $P(\va*{n},t)$ can then be obtained (with some approximations) from Eqs.(\ref{eq:transitionrates}), where $\va*{n} = (n_q^E,n_e^E,n_r^E,n_q^I,n_e^I,n_r^I)$ (See Supplementary Material (SM)). 
As usual \cite{Gardiner}, a Focker-Planck equation can be derived through a Kramers-Moyal expansion, which in turn allows to obtain stochastic Langevin equations for the population densities $\rho_k = \mm{n_k^E/N}$ and $\psi_k = \mm{n_k^I/N}$ ($k=q,\,e,\,r$). Finally, averaging  over the stochastic noise we obtained the  dynamical equations  

\begin{equation}\label{eq:inhibGH}
\begin{split}
    \dot{\rho_e} &= (1-f-\rho_e-\rho_r)\qty(\mu_1 +\alpha \mm{\Theta})- \mu_3\rho_e\\
    \dot{\rho_r} &= \mu_3\rho_e - \mu_2\rho_r\\
    \dot{\psi_e} &= (f-\psi_e-\psi_r)\qty(\mu_1 +\alpha \mm{\Theta}) - \mu_3\psi_e\\
    \dot{\psi_r} &= \mu_3\psi_e - \mu_2\psi_r
\end{split}
\end{equation}
with
\begin{equation*}
    \mm{\Theta} \equiv \Biggl\langle \Theta\qty(\sum_{j=1}^N w_{ij}\epsilon_j\delta(x_j(t),1)-T)\Biggr\rangle
\end{equation*}
\noindent where $\rho_q=1-f-\rho_e-\rho_r$, $\psi_q=f-\psi_e-\psi_r$, and $\mm{\,\cdot\,}$ is an average over the stochastic process $\va*{x}$ and over the quenched disorder ($\{w_{ij},\epsilon_i\}$). Details of the whole procedure to derive Eqs.(\ref{eq:inhibGH}) are presented in the SM. The derivation followed closely similar calculations in a related contact process model in Ref.\cite{corral}.

Assuming that the Heaviside argument behaves as a Gaussian variable we can calculate the average 

\begin{equation}\label{eq:gaussian}
\begin{split}
    \mm{\Theta} &= \frac{1}{\sigma \sqrt{2\pi}}\int_{-\infty}^{\infty} \Theta(v_i) e^{-\frac{(v_i-\mu)^2}{2\sigma^2}}\dd{v_i}\\
    &= \frac{1}{2}\qty[1+\text{erf}\qty(\frac{\mu}{\sqrt{2}\sigma})] \equiv \eta(\mu/\sigma)
\end{split}
\end{equation}
where $\sigma^2 = \text{Var}(v_i)$ and

\begin{equation*}
\begin{split}
    \mu &= \mm{v_i} = \Biggl\langle\sum_{j=1}^N w_{ij}\epsilon_j\delta(x_j(t),1)-T\Biggr\rangle\\
    &\approx \frac{\omega}{N} \Bigg\langle  \sum_{j\in \text{exc}}^{N_{\text{exc}}} \delta(x_j(t),1) - \sum_{j\in \text{inhib}}^{N_{\text{inhib}}} \delta(x_j(t),1)\Bigg\rangle - T\\
    &\approx \omega(\rho_e-\psi_e)-T
\end{split}
\end{equation*}

\noindent where 

\begin{equation}\label{omega}
\omega \equiv \mm{W_{ij}}. 
 \end{equation}
From Eq.(\ref{eq:hagmann}), we get $\omega=1/\beta$. We have also assumed that
\[ \langle  w_{ij}\epsilon_j\delta(x_j(t),1) \rangle\approx \langle  w_{ij}\rangle \langle \epsilon_j \delta(x_j(t),1) \rangle\]

We checked the validity of this approximation by performing numerical simulations of the microscopic GH model in a fully connected network for different values of $T$ and $N$. We found that the correlations between the quenched disorder variables and  $\delta(x_j(t),1)$ were indeed negligible for large enough values of $N$.  Since the results are not expected to depend on the specific form of the sigmoidal function $\eta(x)$ as long as it has the same asymptotic behaviours  for $x\to\pm\infty$, in what follows we replaced $\eta$ in Eq.(\ref{eq:gaussian}) by the simplified form

\begin{equation}
    \eta(x) = \frac{e^{2x}}{1+e^{2x}} \label{eq:eta}
\end{equation}

Replacing the above results into Eqs.(\ref{eq:inhibGH}) we finally obtained

\begin{subequations}
\begin{align}
    \dot{\rho_e} &= (1-f-\rho_e-\rho_r) \,\times \notag\\
    &\times\qty(\mu_1 +\alpha\, \eta\qty(\frac{\omega(\rho_e - \psi_e)-T}{\sigma})) - \mu_3\rho_e\label{eq:inA}\\
    \dot{\rho_r} &= \mu_3\rho_e - \mu_2\rho_r\label{eq:inB}\\
    \dot{\psi_e} &= (f-\psi_e-\psi_r)\, \times \notag\\
    &\times \qty(\mu_1 +\alpha\, \eta\qty(\frac{\omega(\rho_e-\psi_e)-T}{\sigma})) - \mu_3\psi_e\label{eq:inC}\\
    \dot{\psi_r} &= \mu_3\psi_e - \mu_2\psi_r \label{eq:inD}
\end{align}
\label{eq:inhibGH-final}
\end{subequations}

We focused on the stationary solutions of Eqs.(\ref{eq:inA})-(\ref{eq:inD}) (with their corresponding stability) and the dynamical phase transition between them in the $(f,T)$ space. The solutions were obtained analytically, when possible, and numerically using fourth-order Runge-Kutta method when needed.

The model parametrization, {\it i.e.}, the relationship between the rates ($\mu_1$, $\mu_2$, $\mu_3$, $\alpha$) and the microscopic parameters ($r_1$, $r_2$), can be estimated under some general assumptions. Details of the procedure and a validation through numerical comparisons are presented in the Supplementary Material. We found
\begin{equation}\label{eq:relations}
    \mu_1 = \frac{\alpha\, r_1}{1-r_1}\,;\quad \mu_2=\frac{\alpha\, r_2}{1-r_1}\,;\quad \mu_3 = \frac{\alpha}{1-r_1}
\end{equation}
Since all the rates are proportional to $\alpha$, this quantity can be eliminated from Eqs.(\ref{eq:inA})-(\ref{eq:inD}) by a simple re-scaling of the time, so hereafter we set $\alpha=1$.
It is expected that  $\sigma^2 \sim 1/N$. We verified this property by means of numerical simulations of the microscopic model on a fully connected network for different values of $f$ and $T$ (see SM). Hence, in the thermodynamic limit $N\to\infty$ we have that $\sigma\to 0$. We also made some comparisons between stationary values of $\rho_e$ and $\psi_e$ obtained by solving numerically the model with $\sigma \neq 0$ and numerical simulations of the microscopic model on a fully connected network.

\section{Results}\label{results}

We studied the stationary solutions of Eqs.(\ref{eq:inA})-(\ref{eq:inB}) for small but different from zero values of $\sigma$ and then we analysed the thermodynamic limit $\sigma\to 0$.

\subsection{Purely excitatory neurons ($f=0$)}

We first analysed the particular case $f=0$, {\it i.e.}, a system with no inhibitory units. When $f=0$ we have $\psi_q=\psi_e=\psi_r=0$ and Eqs.(\ref{eq:inA})-(\ref{eq:inB}) reduce to 
\begin{equation}\label{eq:MFf0}
\begin{split}
    \dot{\rho_e} &= (1-\rho_e-\rho_r)\qty(\mu_1 +\eta\qty(\frac{\omega\rho_e-T}{\sigma})) - \rho_e\mu_3\\
    \dot{\rho_r} &= \rho_e\mu_3 - \rho_r\mu_2
\end{split}
\end{equation}
First of all, we notice that the stationary solutions of Eqs.(\ref{eq:MFf0}) are constrained to an interval $(\rho_e^{min}, \rho_e^{max})$, where the limiting values are those for which $\eta(..)=0$ and $1$  respectively and are given by

\begin{equation}\label{rhomin}
  \rho_e^{min} =  \, \frac{\mu_1\mu_2}{S}, 
\end{equation}
and
\begin{equation}\label{rhomax}
  \rho_e^{max} =\, \frac{\mu_2(1+\mu_1)}{S+\mu_2+\mu_3},  
    \end{equation}
\noindent  where $S = \mu_1\mu_2 + \mu_2\mu_3 + \mu_3\mu_1$. Notice that these values are independent of $\sigma$. 
Assuming $r_1\ll1$, we get $\rho_e^{min}=\frac{r_1r_2}{r_1+r_2+r_1r_2} \simeq r_1$: the minimum activity 
corresponds to the spontaneous activation solely. Similarly, we get $\rho_e^{max}=\frac{r_2}{2r_2+1}=(2+1/r_2)^{-1}$, as expected\cite{nota}. These results are in complete agreement with previous ones \cite{TesisBarzon,Homeostatic}. 

\begin{figure}
    \centering
    \includegraphics[width=0.96\linewidth]{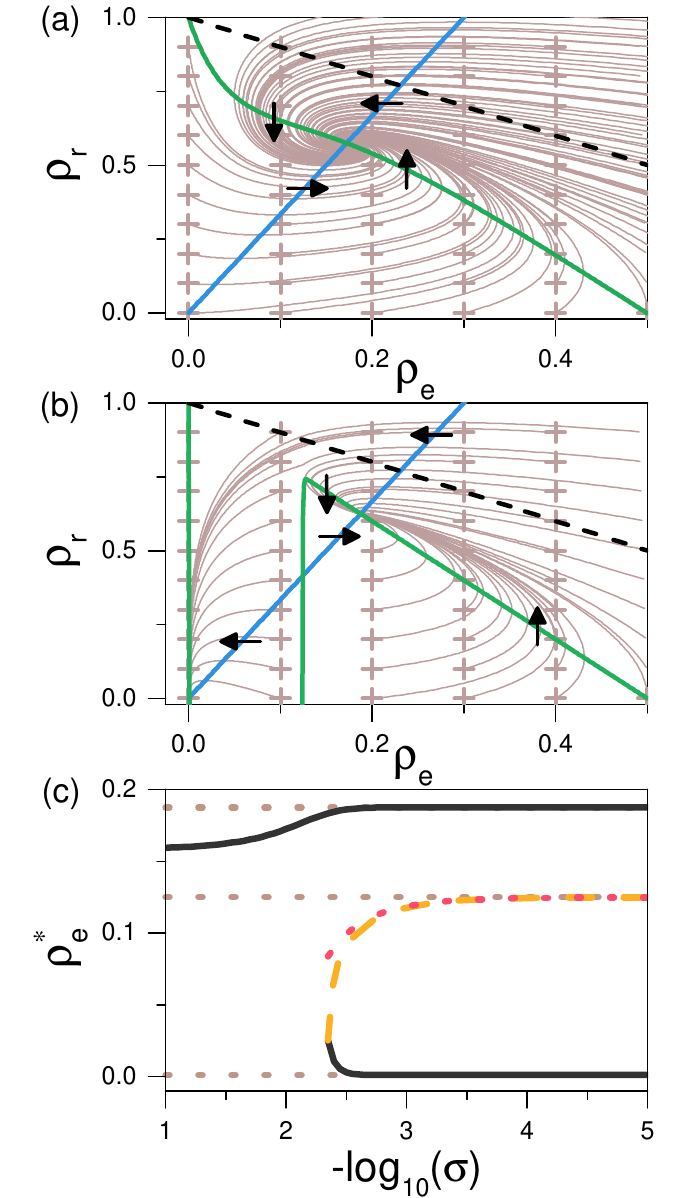}
    \caption{Stationary solutions of the dynamical equations for $f=0$. Panel (a): phase portrait for $T=0.01$   $\sigma=10^{-2}$. Thick blue line corresponds to the nullcline  $\dot{\rho_r}=0$, while the green  line corresponds to $\dot{\rho_e}=0$. Black arrows show the flow directions at the nullclines. Several trajectories starting at the $+$ signs are shown in brown. The black dashed line delimits the area of physically relevant solutions ($\rho_e+\rho_r<1$). Panel (b): same as (a) for $\sigma=10^{-4}$. Panel (c): stationary solutions as a function of $\sigma$. Black lines correspond to stable solutions while orange dashed lines correspond to unstable solutions. The red dashed line corresponds to the polynomial approximation of $\rho^{mid}$ discussed in the SM. From top to bottom, dotted lines correspond to $\rho_e^{max}$, $\frac{T}{\omega}$ and $\rho_e^{min}$}.
    \label{fig:sigma}
\end{figure}

To study the stationary states of Eqs.(\ref{eq:MFf0}) and their stability, we analysed the associated nullclines, {\it i.e.}, the curves defined by the conditions $\dot{\rho_r}=0$ and $\dot{\rho_e}=0$, whose intersections give the corresponding fixed points. The typical behaviour of the nullclines is shown in Fig.\ref{fig:sigma} for different values of $\sigma $. The curve corresponding to $\dot{\rho_r}=0$ is a straight line $\rho_r^*= \rho_e^*\mu_3/\mu_2$, where the asterisk holds for stationary solutions (or long time averages in the numerical model). The curve corresponding to $\dot{\rho_e}=0$ exhibits a non-monotonous behaviour for several values of $\sigma$ and $T$. 

For relatively large values of $\sigma$, and for any value of $T$, we have only one fixed point, which is stable (see Fig.\ref{fig:sigma}-a). As $\sigma$ is decreased, we find several values of $T$ for which there are three fixed points, two stable and one unstable in between (see Fig.\ref{fig:sigma}-b). The new solutions are generated through a perturbed pitchfork bifurcation.

Finally, in the limit $\sigma\to0$, the stable solutions converge to $\rho_e^{min}$  and $\rho_e^{max}$,  given by Eqs.(\ref{rhomin}) and (\ref{rhomax}). The unstable solution $\rho^{mid}$ is close to  $\frac{T}{\omega}$. Indeed, an estimate of its value can be obtained expanding Eqs.(\ref{eq:MFf0}) about $\rho_e=\frac{T}{\omega}$, see the SM.
%Fig. \ref{fig:sigma}-c. 

\begin{figure}
    \centering
    \includegraphics[width=\columnwidth]{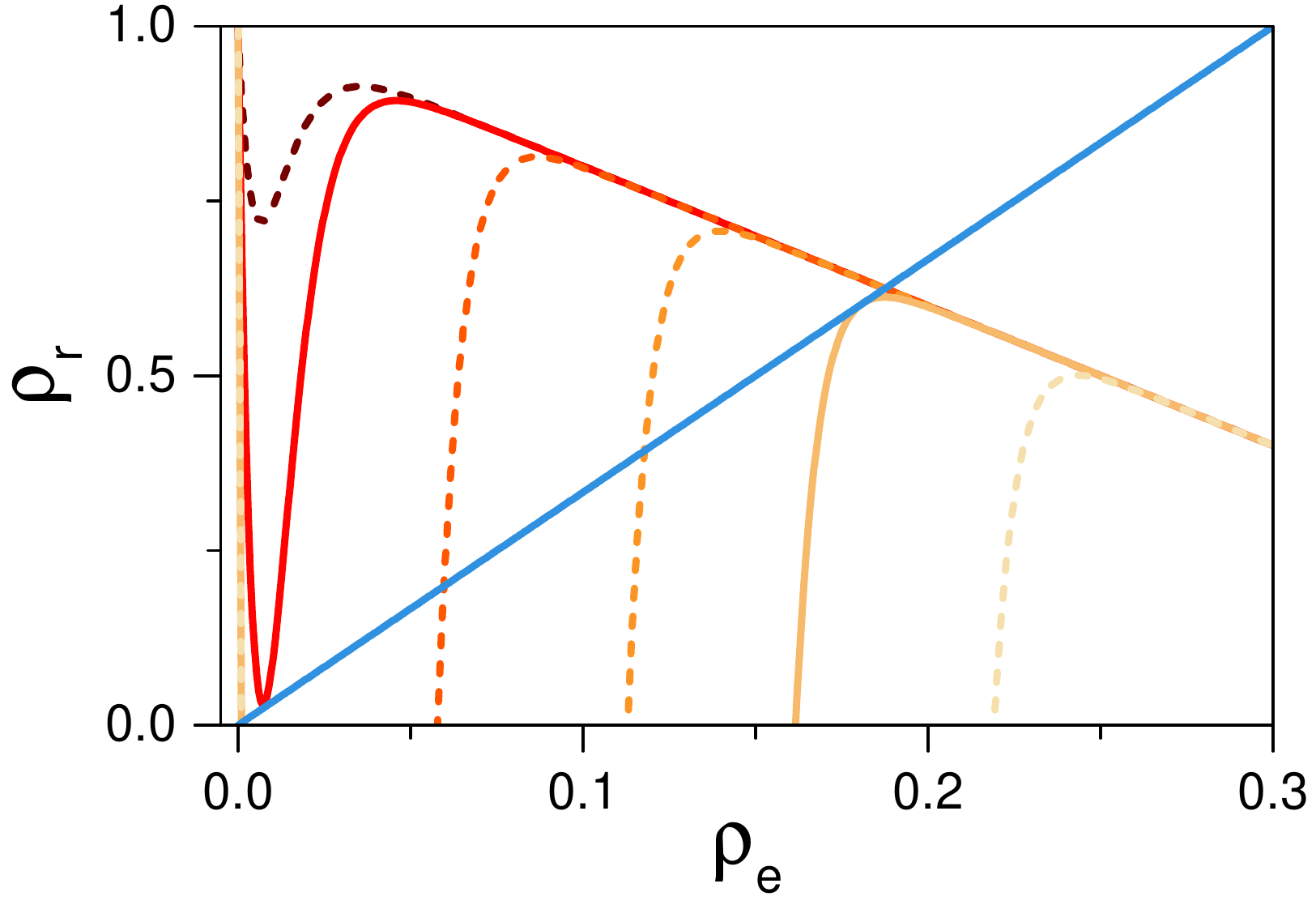}
    \caption{Stationary solutions as a function of $T$ for $f=0$. Nullclines of the dynamical equations for $\sigma=10^{-3}$. The blue line corresponds to $\dot{\rho_r}=0$, while the other lines correspond to $\dot{\rho_e}=0$ for different values of $T$, increasing from left to right. The full line nullclines correspond to the limiting values $(T_{min},T_{max})$  respectively, such that the system presents three fixed points only when $T_{min} < T < T_{max}$, for $\sigma=10^{-3}$.} %Panel (b): Nullcline crossings for very small value of $\sigma$ ($\sigma=10^{-7}$). Black lines are for stable solutions. Orange dashed line is for the unstable solution and the red dotted line is for the line $\beta\cdot T$. Red horizontal lines are for $T_{min/max}^{0}$ and blue vertical lines are for $\rho_{min/max}$  
    \label{fig:nullclinef0}
\end{figure}

The nullclines for fixed $\sigma$ and several values of $T$ are shown in Fig.\ref{fig:nullclinef0}. We observe a range of values  $T_{min}\leq T\leq T_{max}$ for which there are three fixed points. Outside such a range of values of $T$, there is only one stable fixed point. Such behaviour is characteristic of a discontinuous transition and the presence of hysteresis is expected. 

We verified the presence of hysteresis and bistability by performing a loop of increasing-decreasing $T$ and solving Eqs.(\ref{eq:MFf0}) using the Runge-Kutta (RK) method. For each value of $T$, we run the RK algorithm until the system reaches a stationary state and record the stationary value of $\rho_e$ before changing $T \to T\pm \Delta T$ ($\Delta T \sim 1.5\times10^{-5}$) and restart the RK algorithm with the previous state as the new initial condition. We also performed a similar calculation using the cellular automaton defined by Eqs.(\ref{eq:probGH}) in a fully connected network. In this case, for every value of $T$, we discard the first $t=500$ steps and average the fraction of active sites over $10^3-10^4$ steps before changing $T$. In both cases, we observed a clear hysteresis loop. The whole phenomenology can be observed for a wide range of variations of the parameters $(\mu_1,\mu_2,\mu_3,\sigma)$ and has been reported for the same cellular automaton on a Watts-Strogatz network with average degree $20< \langle k \rangle < 40$ and large enough values of the rewiring probability $\pi$\cite{Zarepour,Sanchez}.

\subsection{Excitatory and Inhibitory neurons}
We next considered the general case $f\neq 0$. In this case, we deal with four equations, nevertheless, we can define the following change of variables:

\begin{eqnarray}\label{eq:newvars}
    &\Delta_{e/r}&=f\rho_{e/r}-(1-f)\psi_{e/r} \nonumber \\  &\Sigma_{e/r}&=\rho_{e/r}+\psi_{e/r}
\end{eqnarray}

The evolution equations for $\Delta_{e/r}$  can be written as:

\begin{subequations}
\begin{align}
   \dot{\Delta_e} &= (-\mu_3-\mathcal{N}) \Delta_e  -\mathcal{N} \Delta_r \label{eq:newA}\\
    \dot{\Delta_r} &= \mu_3 \Delta_e  -\mu_2 \Delta_r \label{eq:newB}    
\end{align}
\label{eq:inhibGH-new}
\end{subequations}
%\[
 %\left(\begin{array}{c}
  %       \dot{\Delta_e} \\
   %        \dot{\Delta_R} 
    %     \end{array}
     %    \right)=
%\left(
%  \begin{array}{ccc}
%    -\mu_3-\mathcal{N} & - \mathcal{N} \\
%    \mu_3  & -\mu_2 \label{matrix}  
%  \end{array}
%\right)\cdot
% \left(\begin{array}{c}
 %        \Delta_e \\
  %         \Delta_R 
   %    \end{array} \right), 
%\]

\noindent where $\mathcal{N}\doteq \mu_1 + \eta\qty(\frac{\omega(\rho_e-\psi_e)-T}{\sigma})>0$, and $\rho_e-\psi_e$ can be written as  $(1-2f)\Sigma_e+2\Delta_{e}$. It is straightforward to notice that $\Delta_e= \Delta_r=0$ is a solution of Eqs.(\ref{eq:inhibGH-new}) in the stationary state. Moreover, it can be shown that it is the only solution and that it is locally stable (see SM). This means that $\Delta_{e/r}$ monotonically decay to zero.  Consequently, after a transient, we have $\Delta_{e/r}=0$ and $\psi_{e/r}=\rho_{e/r}\frac{f}{1-f}$, which is reasonable since each inhibitory neuron receives, on average, the same stimulus as each excitatory neuron. We have verified that this decay holds both in the model equations and in the numerical simulations. The equations for $\Sigma_{e/r}$ read:

%The eigenvalues of the matrix are $\lambda_{\pm }={-\mathcal{N}-\mu_3-\mu_2 \over 2} \pm  {1 \over 2} |\mathcal{N}-\mu_3-\mu_2|$, which are both real and negative. This means that $\Delta_{e/r}$ monotonically decay to zero (although the decay exponent may be time dependent since $\mathcal{N}$ is not constant). 

\begin{subequations}
\begin{align}
    \dot{\Sigma_e} &= (1-\Sigma_e-\Sigma_r) \times \notag\\ & \times \small{\qty(\mu_1 +\eta\qty(\frac{\omega [(1-2f) \Sigma_e+2\Delta_e]-T}{\sigma})) - \Sigma_e\mu_3} \label{eq:MFf0edtA}\\
    \dot{\Sigma_r} &= \Sigma_e\mu_3 - \Sigma_r\mu_2.\label{eq:MFf0edtB} 
\end{align} 
\label{eq:MFf0edt}
\end{subequations}

Notice that in the stationary state $\Delta_e=0$, and then, Eqs.(\ref{eq:MFf0edt}) for $\Sigma_{e/r}$ are the same as Eqs.(\ref{eq:MFf0}) for  $\rho_{e/r}$ in the purely excitatory case after rescaling $T\to T/(1-2f)$ and $\sigma \to \sigma/(1-2f)$. We can also rewrite Eqs.(\ref{eq:MFf0edt}) in terms of the excitatory units activity, since $\rho_{e/r}=\Sigma_{e/r}(1-f)$. In particular, Eqs.(\ref{rhomin}) and (\ref{rhomax}) now read:

\begin{equation}\label{rhominnew}
  \Sigma_e^{min} =  \, \frac{\mu_1\mu_2}{S}, 
\end{equation}
and
\begin{equation}\label{rhomaxnew}
  \Sigma_e^{max} =  \, \frac{\mu_2(1+\mu_1)}{S+\mu_2+\mu_3},  
\end{equation}
    
\noindent or, equivalently,  
\begin{align}
    \rho_e^{min} &= (1-f) \, \frac{\mu_1\mu_2}{S} \\
     \rho_e^{max} &= (1-f) \, \frac{\mu_2(1+\mu_1)}{S+\mu_2+\mu_3}.
\end{align}    
    
\begin{figure}
    \centering
    \includegraphics[width=\columnwidth]{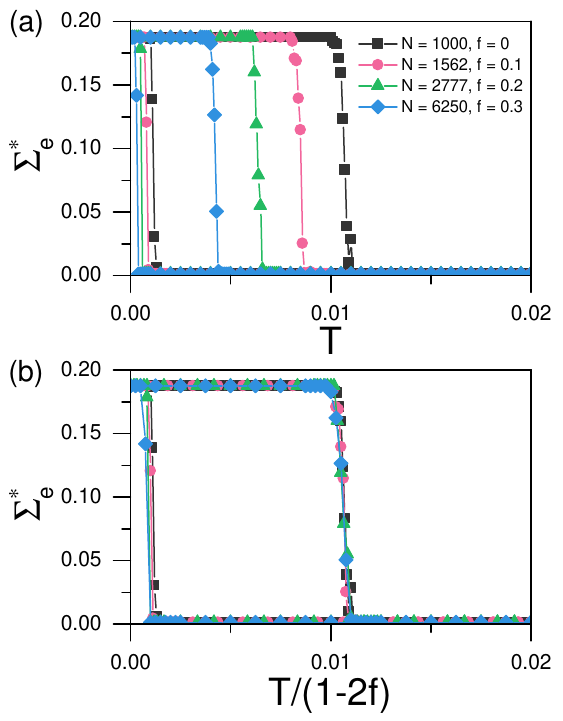}
    \caption{Total activity $\Sigma_e^*$  in numerical simulations, for different values of $f<1/2$. Panel (a): $\Sigma_e^*$  as a function of $T$ for several values of $f$, in systems of size $N=\frac{N_0}{(1-2f)^2}$ with $N_0=1000$. Panel (b): Same data as in panel (a), as a function of $T/(1-2f)$. }
    \label{fig:Equivalence}
\end{figure}

\noindent We verified that these rescaling results also hold for the numerical simulations for finite $N$ (see Fig.\ref{fig:Equivalence}). Since $\sigma^2\sim \frac{1}{N}$ (see SM), we considered systems of size $\frac{N_0}{(1-2f)^2}$ with $N_0=1000$.

For $f\geq 1/2$, Eq.(\ref{eq:MFf0edtA}) predicts that the fraction of active neurons $\Sigma_e$ will decrease as $\sigma$ decreases,  converging to the minimum possible value Eq.(\ref{rhominnew}) in the limit $\sigma\to0$, for any value of $T$. 
  
The stationary solutions of Eqs.(\ref{eq:MFf0edtA})-(\ref{eq:MFf0edtB}) satisfy $\Sigma_r^*=\Sigma_e^*\, \mu_3/\mu_2$  and

\begin{equation}
   T=R(\Sigma_e^*,f) \label{eqTR}
\end{equation}

\noindent  where

%\begin{eqnarray}
\begin{align}
R(\Sigma_e,f) &=
 \frac{\sigma}{2}\ln{\qty[\frac{\mu_2-(\mu_2+\mu_3)\Sigma_e}{S\Sigma_e-\mu_1\mu_2}-1]}+ \notag\\
  &+  \omega(1-2f)\Sigma_e,\label{eq:solorho}
 % \begin{align}
\end{align}
%\end{eqnarray}

\begin{figure}
    \centering
    \includegraphics[width=\columnwidth]{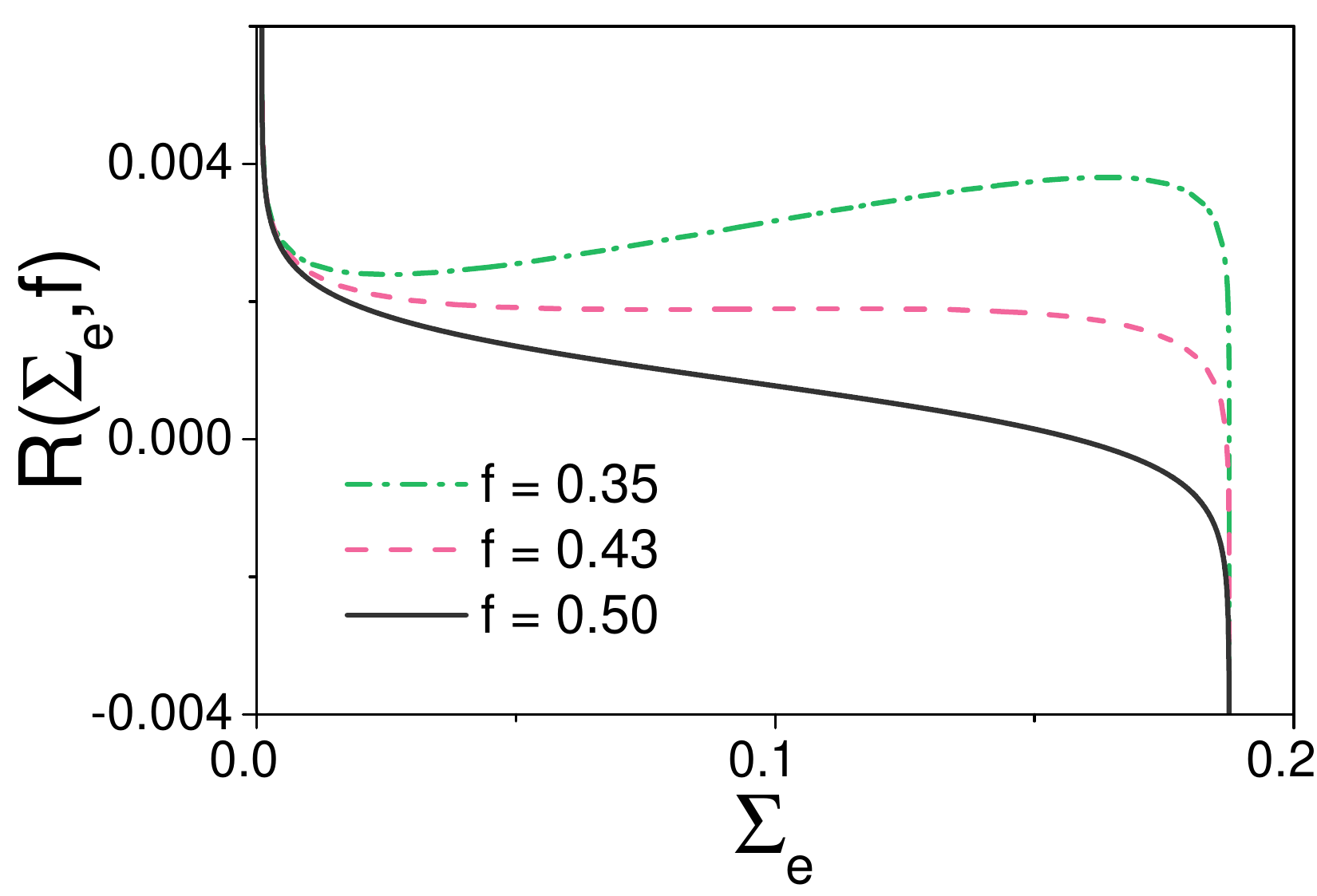}
    \caption{Typical behaviour of  $R(\Sigma_e,f)$ as a function of $\Sigma_e$ for $\sigma=10^{-3}$ and different values of $f$.}
    \label{fig:Rrhoe}
\end{figure}

\begin{figure}
%    \centering
    \includegraphics[width=\columnwidth]{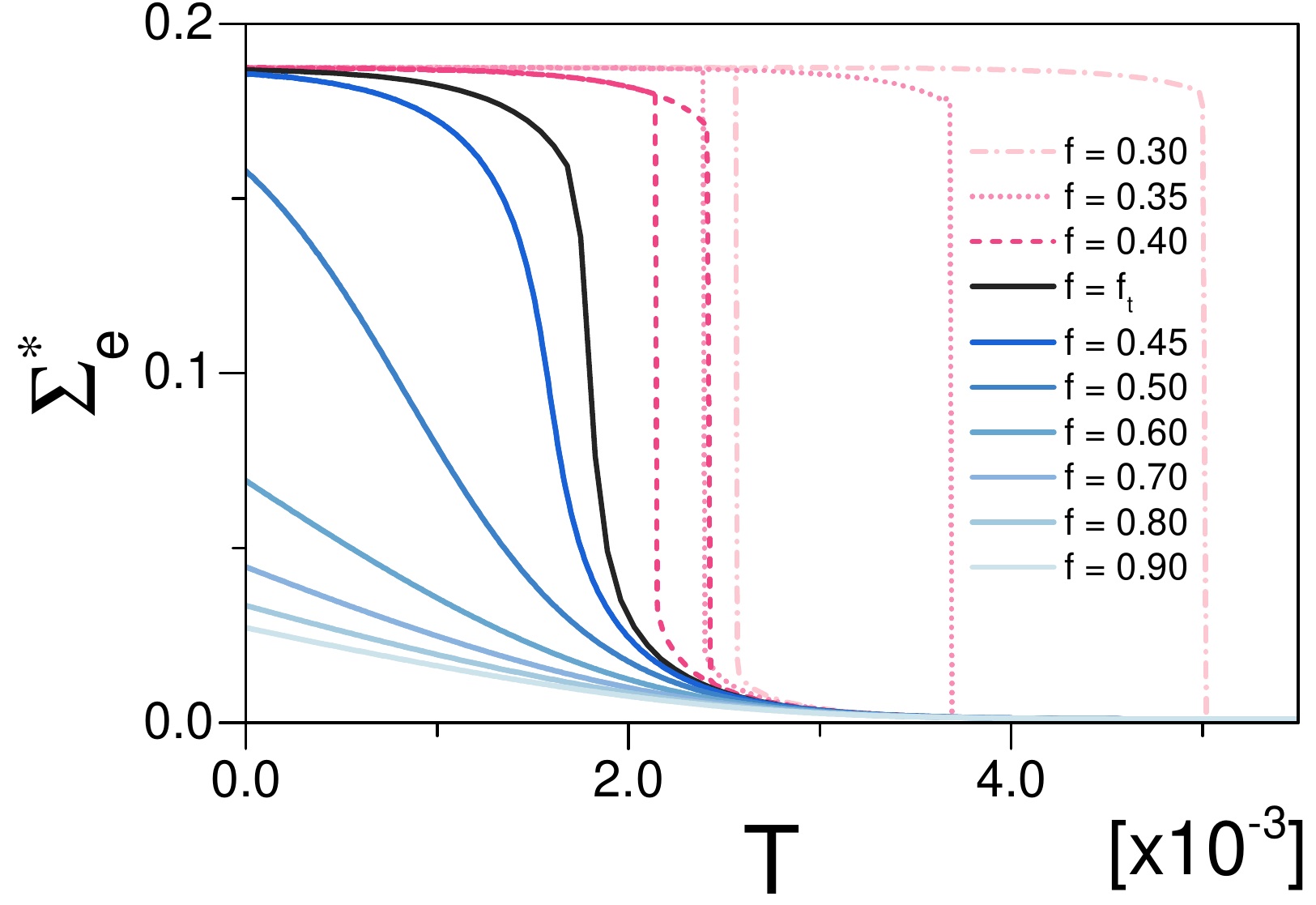}
\caption{Total activity (excitatory and inhibitory neurons) during a cycling process of $T$, obtained from numerical solutions of Eqs.\ref{eq:inhibGH-final} for $\sigma=10^{-3}$ and different values of $f$. A clear hysteresis (red curves) is observed for $f<f_t$. For $f>f_t$ there is no hysteresis and the system exhibits a continuous crossover (solid blue lines) between a low activity regime (for high $T$) to a high activity one (low values of $T$).}
    \label{fig:orederparmodel}
\end{figure}

From Eq.(\ref{eq:solorho}), we find that the bistability limits in the thermodynamic limit ($\sigma\to0$ in the model, or $N\to\infty$ in the numerical simulations) are: 
%\noindent $\omega$ is given by Eq.(\ref{omega}).
\begin{equation}\label{eq:Tlimits}
T_{min/max}=\omega(1-2f)\Sigma_e^{min/max},
\end{equation}
\noindent having then a finite range of $T$ values where there is bistability.
For finite $N$, we find that $T_{min/max}$ converge rather slowly (as a power law) to the values given by Eq.(\ref{eq:Tlimits}) (see SM).
The typical behaviour of  $R(\Sigma_e,f)$ as a function of $\Sigma_e$ is shown in Fig.\ref{fig:Rrhoe}.  When $\sigma \ll 1$, the behaviour of $T$ as a function of $\Sigma_e$ is dominated by the linear term proportional to $\omega$, except for values of $\rho_e$ very close to $\Sigma_e^{min}$ or $\Sigma_e^{max}$. Since $\omega (1-2f)$ changes of sign at $f=1/2$, we see that there exists a value $f_t\approx 1/2$, such that when $f < f_t$, $R(\Sigma_e,f)$ has a positive slope and Eq.(\ref{eqTR}) has three solutions that converge to the previous analysed ones (i.e., when $f\to 0$). Hence, one could expect a discontinuous transition between low and high activity phases and hysteresis in the whole range $f<f_t$. On the contrary, when $f\geq f_t$, there is only one solution that changes continuously from high values of the activity for small values of $T$ to small activities for high values of it. Numerical solutions of Eqs.(\ref{eq:inA})-(\ref{eq:inD}) analogue to those performed for $f=0$ ({\it i.e.}, cycling $T$) support the above conclusions. The corresponding results for the stationary values of the total activity $\Sigma_e^*=\rho_e^*+\psi_e^*$ as a function of $T$ for different values of $f$ and a finite but small value of $\sigma$ are shown in Fig.\ref{fig:orederparmodel}. The behaviour of the activity when $f>f_t$ is consistent with the expected finite size behaviour of a second order phase transition order parameter, at least for values of $f$ close to $f_t$. In that sense, the end point of the discontinuous transition line $(f,T)=(f_t,T_t)$ could be regarded as a tricritical one. However, we will show that such pseudo critical behaviour for $f>f_t$ completely disappears in the thermodynamic limit $\sigma\to 0$.

The pseudo tricritical value $f_t$ corresponds to an inflexion point of $R(\Sigma_e,f)$. So, equating the first and second derivatives of Eq.(\ref{eq:solorho}) to zero and solving together with Eq.(\ref{eq:solorho}) we obtain after some algebra 

\begin{equation}
 \Sigma_e^t =    \frac{\mu_1\mu_2(S+\mu_2+\mu_3)+S(\mu_2+\mu_1\mu_2)}{2S(S+\mu_2+\mu_3)} %\\ 
 % &=\frac{\mu_1\mu_2}{2S}+\mu_2\frac{\mu_1+1} {2(S+\mu_2+\mu_3)} \nonumber 
\end{equation}

\begin{equation}
    f_t = \frac{1}{2}-\frac{\sigma}{4\omega}B
\end{equation}

\begin{equation}
    T_t = \frac{\sigma}{2}\qty(\Sigma_e^t B + \ln{\qty(\frac{\mu_2-(\mu_2+\mu_3) \Sigma_e^t}{S \Sigma_e^t -\mu_1\mu_2}-1)})
\end{equation}

\noindent where

\begin{equation*}
    B = \frac{4S(S+\mu_2+\mu_3)}{\mu_2^2\mu_3}
\end{equation*}

\begin{figure}
    \centering
    \includegraphics[width=\columnwidth]{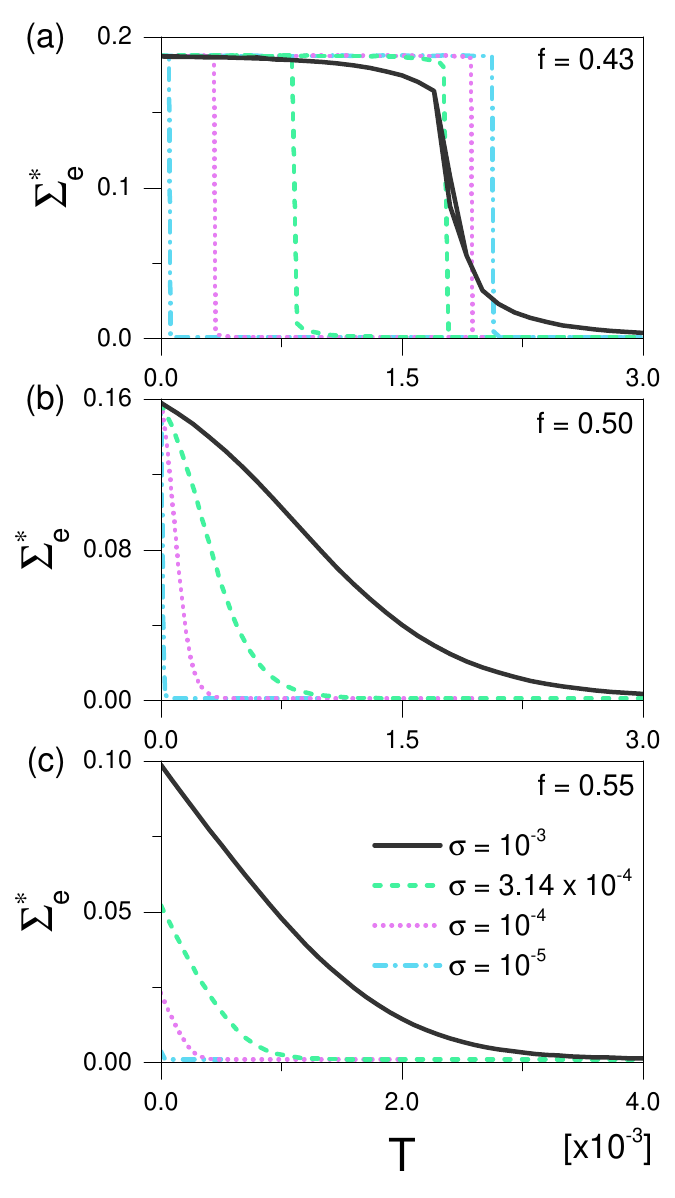}
    \caption{Total activity during a cycling process of $T$, obtained from  numerical solutions of Eqs.(\ref{eq:inhibGH-final}) when $\sigma\to 0$ for values of $f$ around $1/2$.}
    \label{fig:ordparvssig}
\end{figure}

\begin{figure}
    \centering
    \includegraphics[width=\columnwidth]{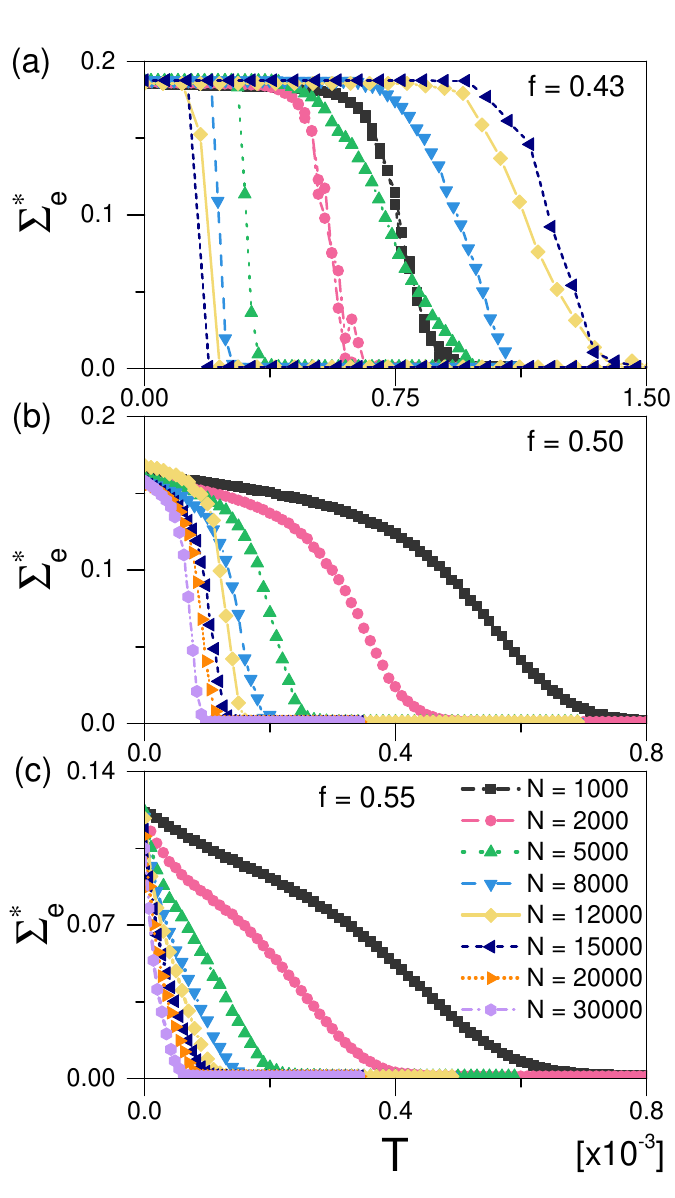}
    \caption{Average total activity during a cycling process of $T$, obtained from numerical simulations of the microscopic model defined by Eqs.(\ref{eq:probGH}) in a fully connected network of $N$ sites when $N$ increases.   When $f<1/2$ we observe the appearance of a hysteresis cycle for large enough system sizes, that converge to a well defined one in the limit $N\to\infty$. Hysteresis disappears for $f\geq 1/2$ for any value of $N$.  }
    \label{fig:size-scaling}
\end{figure}

In the thermodynamic limit $\sigma\to 0$ we see that $f_t \to 1/2$ and $T_t\to 0$. To understand the last result we calculated again the numerical solutions of the dynamical equations in cycling process of $T$, for a sequence of decreasing values of $\sigma$ and values of $f$ around $1/2$. The main results of these calculations are shown in Fig.\ref{fig:ordparvssig}. We see that in the limit $\sigma\to 0$ the global activity becomes negligible ($\sim \rho_e^{min} \sim 0.5 \times 10^{-3}$ in the present case) for any value of $T>0$ when $f\geq 1/2$, while a well defined hysteresis loop is established when $f<1/2$. Consistently, we obtained the same results when simulating the microscopic model in a fully connected network, following the same cycling protocol for increasing values of the system size $N$ (every curve was averaged in this case over the quenched disorder in both the synaptic weights and the distribution of inhibitory units), as shown in Fig.\ref{fig:size-scaling}. Extrapolation of these curves to $N\to\infty$ (not shown) when $f\geq 1/2$ confirms that the activity becomes negligible for any value of $T>0$. 

%\textcolor{red}{This result is not enterely surprising since for $f>1/2$ there are more inhibitory than excitatory neurons, and the intensity of interactions (given by the $W_{ij}$  matrix) on average, is the same no matter the nature of the presynaptic (i.e., $j$) or postsynaptic  (i.e., $i$) neurons. Then any number of excitatory active neurons at time $t$ will activate (on average) more inhibitory than excitatory units at time $t+1$, which will make the total activity disappear at time $t+2$, since each postsynaptic neuron will receive more negative than positive inputs. Then, the only possible activity for $f>1/2$ is due to the finite size fluctuations in the number of active excitatory and inhibitory units, which is only relevant for small $N$ }

This result can be understood in the following way. Since for $f>1/2$ there are more inhibitory than excitatory neurons, in a fully connected network any number of excitatory active neurons at time $t$ will activate (on average) more inhibitory than excitatory units at time $t+1$, which will make the total activity decrease at time $t+2$. Hence, global activity is expected to decrease monotonically in the long term and the only possible activity for $f>1/2$ in the stationary state is due to the finite size fluctuations in the number of active excitatory and inhibitory units, which is only relevant for small $N$.

\section{Discussion}\label{conclusion}

We obtained the mean field dynamical equations of a Greenberg-Hastings neural model with excitatory and inhibitory units. The dynamical equations results and the numerical simulation of model show a very good agreement. The analysis of the stationary solutions revealed the existence of a first order ({\it i.e.}, discontinuous) dynamical phase transition between an active phase  (high values of the total activity $\Sigma_e^*$) 
 and an inactive one (low values of the total activity $\Sigma_e^*$), where the threshold $T$ plays the role of the control parameter of the transition. Such transition happens for values of the inhibitory neurons fraction $f$ smaller than certain value $f_t \leq 1/2$, even for finite systems and remains in the thermodynamic limit $\sigma\to 0$, where $f_t\to 1/2$.
 
 We found that, after a transient, the number of active inhibitory and excitatory neurons is proportional. Then, the net effect of the inhibitory units is to rescale the value of $T$. This is a consequence of having an interaction matrix $W$ whose values do not depend on the nature (i.e., either inhibitory or excitatory) of the presynaptic and postsynaptic neurons. Notice that other models in the context of excitatory-inhibitory balance (see e.g. \cite{corral} and references therein),  consider similar evolution equations  to those studied here for the fraction of active excitatory and inhibitory neurons (although equations for the refractary units are not  present). In these models, they propose different interaction strengths for the excitatory-excitatory, excitatory-inhibitory, inhibitory-excitatory or inhibitory-inhibitory connections, which allows to  show a wider variety of dynamics \cite{corral}. An interesting avenue of future research is to study under which conditions the proportionality among active excitatory and active inhibitory units  is mantained. 
 
We found a pseudo triciritical point at $f_t\simeq 1/2$, which is quite away from biologically relevant fractions (which are about $f=0.2$). The obtained value of $f_t$ is not relevant on it's own, but rather,  a consequence of having the same strength for inhibitory and excitatory output connections: if we multiply the strength of the elements $W_{ij}$, where $j$ is inhibitory, by a factor $\alpha$, one could expect $f_t$ to be displaced towards $f_t=1/(\alpha+1)$.

From a statistical physics point of view, the presence of a phase transition ({\it i.e.}, a non-analytic behaviour of the state variables)  in a finite system, is not unusual in non--thermodynamical systems \cite{Almeira,Serra, Leviatan}, at variance with what happens in thermodynamical ones where non-analytic behaviour can only happen in the thermodynamic limit. The present example is consistent with an early prediction from a phenomenological model of excitatory and inhibitory neural populations \cite{Wilson}. Nevertheless, we found that the first order transition line in the $(f,T)$ space stops at a point $(f_t,T_t)$, being replaced by a continuous crossover between low and high activity regions when $f>f_t$, that resembles the finite size behaviour of a continuous phase transition order parameter. In that sense, $(f_t,T_t)$ could be regarded as a pseudo tricritical point. However, such pseudo critical behaviour for $f>f_t$  completely disappears in the thermodynamic limit $\sigma\to 0$, where the total activity becomes negligible for any value of $T>0$. On the other hand, the behaviour of the GH model on a fully connected network is expected to reproduce  (with the interactions strength properly normalized) the $\langle k \rangle \to N \to\infty$ limit of the GH model on a Watts-Strogatz network which, for an average degree $\langle k \rangle = 30$  exhibits true tricritical behaviour \cite{Almeira}. Therefore, the present results suggest that the critical region of the last model in the large connectivity limit should shrink as the average degree increases and the tricritical fraction of inhibitory units should converge to 1/2. 
Moreover, preliminary simulation results indicate that the critical region remains finite in large $N\gg 1$  systems with large connectivity  $\langle k \rangle\gg 1$, as long as $\langle k \rangle / N \ll 1$. Works along these lines are in progress and will be published elsewhere.

%%%%%%%%%%%%%%%%%%%%%% ACKNOWLEDGEMENTS %%%%%%%%%%%%%%%%%%%%%%%%%%%%%%%%%%%%%%%

\begin{acknowledgements}
This work was partially supported by CONICET (Argentina) through Grants PIP 
No. 1122020010106, by SeCyT (Universidad Nacional de
Córdoba, Argentina) and by the NIH (USA) Grant No.
1U19NS107464-01. J.A. supported by a Doctoral Fellowship
from CONICET (Argentina). This work used Mendieta Cluster from CCAD-UNC, which is part of SNCAD-MinCyT,
Argentina.
\end{acknowledgements}

%%%%%%%%%%%%%%%%%%%%%% BIBLIOGRAPHY %%%%%%%%%%%%%%%%%%%%%%%%%%%%%%%%%%%%%%%


\begin{thebibliography}{99}
\bibitem{GH} J. M. Greenberg, S. P. Hastings, Spatial Patterns for Discrete Models of Diffusion in Excitable Media, \textit{SIAM J. Appl. Math.},34, 515, (1978).

\bibitem{Haimovici} A. Haimovici, E. Tagliazucchi, P. Balenzuela, D. R. Chialvo, Brain Organization into Resting State Networks Emerges at Criticality on a Model of the Human Connectome, \textit{Phys. Rev. Lett.}, 110, 178101, (2013).
\bibitem{Zarepour} M. Zarepour, J. I. Perotti, O. V. Billoni, D. R. Chialvo, S. A. Cannas, Universal and nonuniversal neural dynamics on small world connectomes: A finite-size scaling analysis, \textit{Phys. Rev. E}, 100, 052138, (2019).
\bibitem{Sanchez} M. M. Sánchez Díaz, E. J. Aguilar Trejo, D. A.
Martin, S. A. Cannas, T. S. Grigera, D. R. Chialvo, Similar local neuronal dynamics may lead to different collective behavior, \textit{Phys. Rev. E}, 104, 064309, (2021).
\bibitem{Almeira} J. Almeira, T. S. Grigera, D. R. Chialvo, S. A. Cannas, Tricritical behavior in a neural model with excitatory and inhibitory units, \textit{Phys. Rev. E}, 106, 054140, (2022).
\bibitem{Beggs}  J. M. Beggs and D. Plenz, Neuronal avalanches in neocortical circuits, \textit{J. Neurosci.}, 23, 11167 (2003).
\bibitem{Chialvo} D. R. Chialvo, Emergent complex neural dynamics, \textit{Nat. Phys.}, 6, 744 (2010).
\bibitem{Mora}  T. Mora and W. Bialek, Are biological systems poised at criticality?, \textit{J. Stat. Phys.}, 144, 268 (2011).
\bibitem{Hagmann} P. Hagmann, L. Cammoun, X. Gigandet, R. Meuli, C. J. Honey, V. J. Wedeen, O. Sporns, Mapping the Structural Core of Human Cerebral Cortex, \textit{PLoS Biol.}, 6, e159, (2008).
\bibitem{corral} R. Corral López, V. Buendía, M. A. Muñoz, The excitatory-inhibitory branching process: a parsimonious view of cortical asynchronous states, excitability, and criticality, \textit{Phys. Rev. Res.}, 4, 4, L042027, (2022).
\bibitem{Gardiner} C. Gardiner, \textit{Stochastic Methods, A Handbook for the Natural and Social Sciences} (Springer Series in Synergetics), Springer Berlin, Heidelberg, (2009).
\bibitem{nota} This last result can be easily understood if each single neuron fires at its maximum rate: it will spend 1 step in the excited state, on average, $1/r_2$ steps in the refractary state, and after that, just one step in the quiescent state before starting the cycle again. So, the cycle lasts on average $2+1/r_2$ steps, and the neuron spends 1 step per cycle (i.e., $\frac{1}{2+1/r_2}$ of the time) in the excited state.
\bibitem{TesisBarzon} G. Barzon \textit{Structure-function relation in a stochastic whole-brain model at criticality}. Final Dissertation for the Master Degree in Physics of Data, Padova University (2021). https://hdl.handle.net/20.500.12608/21199
\bibitem{Homeostatic} R. P. Rocha, L. Koçillari, S. Suweis \textit{et al.}, Homeostatic plasticity and emergence of functional networks in a whole-brain model at criticality, \textit{Sci. Rep.}, 8, 15682 (2018). \bibitem{Wilson} H. R. Wilson and J. D. Cowan, Excitatory and inhibitory interactions in localized populations of model neurons, \textit{Biophys J.}, 12, (1972).
\bibitem{Serra} P. Serra and J. M. Stilck, Polymer model with annealed dilution on the square lattice: a transfer matrix study, \textit{Phys. Rev. E}, 49, 1336 (1994).
\bibitem{Leviatan} A. Leviatan, First-Order Quantum Phase Transition in a Finite System, \textit{Phys. Rev. C}, 74, 051301 (2006).



\end{thebibliography}
\end{document}